\documentstyle[preprint,aps]{revtex}

\preprint{\bf PREPRINT}

\begin{document}
\columnsep0.1truecm
\draft
\title{Hysteresis Loop Critical Exponents in $6-\epsilon$ Dimensions}
\author{Karin Dahmen and James P. Sethna}
\address{Laboratory of Atomic and Solid State Physics,\\
Cornell University, Ithaca, NY, 14853-2501}
\maketitle

\begin{abstract}
The hysteresis loop in the zero--temperature random--field Ising model
exhibits a critical point as the width of the disorder increases.  Above
six dimensions, the critical exponents of this transition, where the
``infinite avalanche'' first disappears, are described by mean--field theory.
We expand the critical exponents about mean--field theory, in $6-\epsilon$
dimensions, to first order in $\epsilon$.  Despite $\epsilon=3$, the
values obtained agree reasonably well with the numerical values in three
dimensions.

\end{abstract}


\narrowtext
In a previous paper\cite{hyster}, we modeled hysteresis in magnetic and
martensitic systems using the random--field Ising model at zero
temperature.  The model exhibited two features characteristic of these
systems: the return--point memory effect and avalanche--generated noise.
(The noise is called Barkhausen noise in magnetic systems and
acoustic emission
in martensites.)  We also discovered a critical point, separating
smooth hysteresis loops at large disorder where all avalanches are
finite, from discontinuous hysteresis loops at small disorder where one
avalanche turns over a fraction of the whole system.

Here we study this critical point in an expansion about mean--field
theory.  Figure~\ref{flows}(a) shows a schematic of the
phase diagram for our model defined by eq.~(\ref{model}) below.
The vertical axis $H$ is the external
field. The horizontal axis $R$ is the width of the
probability distribution of the random fields $f_i$
acting on each spin.  The bold line represents the location $H_c(R)$
at which the
infinite avalanche occurs, when the field $H(t)$ is adiabatically
increasing from an initial state where all spins were pointing down.  At
small disorder, the first spin to flip easily pushes over its neighbors,
and the transition happens in one burst (the infinite avalanche).  At
large enough disorder, the coupling between spins becomes negligible,
and most spins flip by themselves: no infinite avalanche occurs.  At a
special value of the randomness $R=R_c$ the infinite avalanche
disappears.  We find a critical point with two relevant variables
$r\equiv (R_c-R)/R_c$ and $h \equiv (H-H_c(R_c))$ \cite{hyster}.  At
this point we find a universal scaling law for the magnetization $m
\equiv (M-M_c(R_c))$
\begin{equation}
\label{magnetization}
m \sim |r|^{\beta} {\cal M}_{\pm}(h/|r|^{\beta \delta}) \, ,
\end{equation}
where the $\pm$ refers to the sign of r.

We use a soft-spin
version of the random field Ising model, whose energy at a given
spin configuration $\{s_i\}$ is
\begin{equation}
\label{model}
{\cal H}=-\sum_{ij}J_{ij}s_i s_j-\sum_i f_i s_i+H s_i-V(s_i) \, ,
\end{equation}
with the linear cusp-potential $V(s_i)$ defined\cite{cusp} through
\vspace{.3cm}
\[ V(s_i) = \left\{ \begin{array}{ll}
                            k/2~(s_i+1)^2 & \mbox{for $s<0$} \\
                            k/2~(s_i-1)^2 & \mbox{for $s>0$}
                            \end{array}
                   \right. \]
Here, $k>0$ is the local curvature of the potential.
The spins are coupled ferromagnetically by a nearest neighbor interaction
$J_{ij}=J/z>0$, $z$ being the coordination number of the lattice.
We demand that $k/J>1$ to ensure stability of the
system. $H$ is a homogeneous external
magnetic field; the $f_i$ are randomly chosen from a Gaussian distribution
$\rho(f)$ of standard deviation $R$.
We study this system at zero temperature.  It turns out for given $H$ and
$f_i$ that there are many metastable states; which one of these
the system picks
depends entirely on its history (i.e. the way the external magnetic
field $H$ was varied at earlier times).  We will study the history of a
monotonically but adiabatically increasing external magnetic field. We
impose purely relaxational dynamics, as defined by the equation of
motion
\begin{equation}
\label{equ.of.motion}
\partial_t s_i(t) = - \frac{\delta({\cal H})}{\delta s_i(t)}\, ,
\end{equation}
where we have absorbed the friction constant into the definition of
the time $t$.
\paragraph{Formalism:}
We use the formalism of Martin, Siggia and Rose\cite{MSR} to rewrite the
problem as a path integral for a generating functional $Z$, and then
expand this functional about mean field theory.  This is done in analogy
with the calculation for CDWs by Narayan and
Fisher\cite{Narayan1,Narayan2}. We impose the dynamical
equation~(\ref{equ.of.motion}) on the path integral at each time $t$
by introducing it as a $\delta$-function constraint using the well--known
identity
$2 \pi \delta(f(s)) = \int_{-\infty}^\infty e^{i \hat s f(s)} d\hat s$:
\begin{equation}
 Z \equiv \int [ds] [d\hat{s}] J[s] \exp(S)
\end{equation}
where
\begin{equation}S = \frac{i}{J} \int dt \sum_{j} \hat{s_j}(t)
(\partial_t
s_j(t) - \sum_{\ell} J_{j\ell} s_\ell - H - f_j +
\frac{\delta V}{\delta s_j}) \end{equation}
Derivatives of $Z$ can be used to calculate the response functions and
dynamic correlation functions for our model. $J[s]$ is a functional
Jacobian, chosen such that $Z$ is unity, independent of the $f_j$.  This
allows us to average over the disorder without fancy tricks (like
replica theory).

We choose a particular regularization for the time integral. The
simplest choice\cite{Narayan1} is to require a force at time $t$ to
have an effect only after some time $\delta t$. That leaves us with $J[s]
\equiv 1$.  We now do an average over the random fields $f_i$, denoted
by $\langle \, \rangle_f$, leading to the averaged generating functional
\begin{equation}\bar{Z} = \int [ds] [d\hat{s}] \langle\exp(S)\rangle_f.
\end{equation}

To expand about mean--field theory, we need
change variables from
$s_j$ and ${\hat s}_j$ to the local fields $\eta_j = (1/J)\sum_\ell
J_{j\ell} s_\ell$
at the sites (fluctuations about whose mean values we shall study).
We do so by introducing another auxiliary field $\hat \eta_j$,
and absorb a factor $i$ in its definition, so
\begin{equation}
\label{zbar}
\bar{Z} = \int [d\eta] [d\hat\eta] \prod_j \bar Z_j[\eta_j,{\hat \eta}_j]
		\exp\left\{-\int dt \sum_j {\hat\eta}_j(t)
		\left(\sum_\ell J_{j\ell}^{-1} J\eta_\ell(t)\right) \right\}
\, ,
\end{equation}
where $\bar{Z}_j[\eta_j,\hat{\eta_j}]$ is a local functional
\begin{equation}
\bar{Z}_j[\eta_j,\hat{\eta_j}] = \int [ds] [d\hat{s}]\langle\exp S_j\rangle_f
\, ,
\end{equation}
and
\begin{equation}
S_j=\frac{1}{J} \int dt \left\{\sum_{j} J\hat{\eta}_j(t) s_j(t) + i
 \hat{s_j}(t)\left(\partial_ts_j(t) - J\eta_j
- H - f_j + \frac{\delta V}{\delta s_j}\right)\right\} \, .
\end{equation}
We will now expand about the mean--field solution $\eta_0$
(the local field configuration about which the log of the
integrand in equation~(\ref{zbar}) is stationary).
Shifting the definition of $\eta$ to $\eta -\eta_0$ so that
$\langle\eta\rangle_f = 0$
leaves one
with the generating functional
\begin{equation}
\bar{Z} = \int [d\eta] [d\hat\eta] \exp(\tilde{S})
\end{equation}
with an effective action
\begin{eqnarray}
\tilde{S} &=&
- \sum_{j,l} \int dt J_{jl}^{-1} J\hat{\eta}_j(t) \eta_l(t)
+\sum_j \sum_{m,n=0}^{\infty} \frac{1}{m! n!} \int dt_1\cdot \cdot \cdot
dt_{m+n} \\
& & u_{mn}(t_1,...,t_{m+n})
\hat{\eta}_j(t_1)\cdot \cdot \cdot \hat{\eta }_j(t_m) \eta_j(t_{m+1}) \cdot
\cdot \cdot \eta_j(t_{m+n}) \, .
\nonumber
\end{eqnarray}
Here, the $u_{mn}$ are the derivatives of $\ln \bar{Z}_i$ with
respect to the fields $\hat{\eta}_j$ and $\eta_j$, and thus are equal to
the local, connected responses and correlations in mean-field theory:
\begin{equation}
\label{u_mn}
u_{m,n}= \frac{\partial}{\partial \epsilon(t_{m+1})} \cdot \cdot \cdot
\frac{\partial}{\partial \epsilon(t_{m+n})} \langle s(t_1) \cdot \cdot \cdot
s(t_m)\rangle_{f,l,c}.
\end{equation}
Local\cite{Narayan1} ($l$) means that we do not vary the local field
$(\eta_0)_j$ in the mean--field equation
\begin{equation}
\label{response}
\partial_t s_j(t) = J (\eta_0)_j(t)+H+f_j-
\frac{\delta V}{\delta s_j(t)}+ J \epsilon (t)
\end{equation}
when we perturb with the infinitesimal force $J \epsilon(t)$.
The force $J\epsilon(t)$ is only allowed to increase
with time, consistent with the history we have chosen. (For example
for $u_{1,1}(t,t^{\prime})$ we
add a force $J\epsilon\,\Theta(t-t_{\prime})$ in
eq.~(\ref{response}), with $\Theta(t-t_{\prime})$ being the
step-function,
and take the derivative of $\langle s(t)\rangle_{f,l,c}$ with
respect to $\epsilon$ and $t_{\prime}$.)
\paragraph{RG treatment:}
We consider the $\hat{\eta} \eta$ term in the action (involving
$J_{jl}^{-1}$ and $u_{1,1}$) as the propagator in the RG treatment. Now
we take some long--wavelength and low--frequency limits in analogy
to\cite{Narayan1,Narayan2}.  For small wave vectors $J^{-1}(q) \sim
(1/J) + J_2q^2$, and we rescale to give $J J_2=1$.
We take the low--frequency part of the propagator, by
Fourier transforming the $\hat{\eta } \eta$ term in
time, expanding to first order in $\omega$ and Fourier transforming
back.  The propagator (the $\hat{\eta} \eta$ term in the action) is thus
(after rescaling)
\begin{equation}
 -\int d^dq \int dt \hat{\eta}(-q,t)
	[-\partial_t + q^2 - \chi^{-1}/J]\eta(q,t) \, .
\end{equation}
The bare value of $\chi$ is the static response, calculated in
mean-field theory, to a monotonically
increasing external magnetic field
\begin{equation}
\chi = 1/\left(2 J^2\rho(-JM-H+k)-J(k-J)/k\right) \, ,
\end{equation}
where $M$ is the magnetization at the external magnetic field $H$.

We use the Wilson-Fisher renormalization group transformation: In each
step we integrate out modes of all frequencies and wave vectors within an
infinitesimal wave vector shell\cite{Narayan1}. We rescale through $x=b
x^{\prime}$, $t=b^zt^{\prime}$. We choose the rescaling of the fields
such that the $q^2$ term of the propagator and the $u_{2,0}$ term remain
unchanged, since to first order in $\epsilon$ they have no loop
corrections. Thus $\hat{\eta} = b^{-\frac{d}{2}-z} {\hat{\eta}}^{\prime}$
and $\eta =b^{-\frac{d}{2}+2} {\eta}^{\prime}$\cite{Narayan2}.
Without loop corrections this implies $z=2$.
Keeping in mind that the $\partial/\partial \epsilon(t)$ in
eq.~(\ref{u_mn}) rescale like $b^{-z}$, we arrive at
$u_{mn}' = b^{[-(m+n)+2]d/2+2n}u_{mn}.$
To lowest order in $\epsilon=d_c-d$, the only relevant terms are those which
do not flow to zero under rescaling at the upper critical dimension
$d_c$ (we will see that $d_c=6$ for our critical point).
$u_{1,0}$ is trivially zero because we
expand around the stationary-point. The $u_{1,2}$ term, in the static limit,
has bare value $w= -2J^2\rho^{\prime}(-JM-H+k)$: it becomes relevant for
$d<8$.  The $u_{1,3}$ term starts at $u=2J^3\rho^{\prime \prime}(-JM-H+k)$ in
the static limit, and is relevant for $d<6$.  Finally, the
$u_{2,0}$ term stays marginal. In the static limit that we
consider $u_{2,0}$ can be treated as a constant.
We are left with the effective action
\begin{eqnarray}
\tilde S &=& -\int d^dq \int dt\, \hat{\eta}(-q,t)
	[-\partial_t + q^2 - \chi^{-1}/J]\, \eta(q,t)
+ \sum_{j} \Bigl[\,(1/2) \int dt\, \hat{\eta}_j(t) (\eta_j(t))^2 w
\nonumber \\
& & + (1/6) \int dt\, \hat{\eta}_j(t) (\eta_j(t))^3 u
+ (1/2) \int dt_1 \int dt_2\, \hat{\eta}_j(t_1) \hat{\eta}_j(t_2)
u_{2,0}(t_1,t_2)\Bigr]
\end{eqnarray}

Our $\epsilon$-expansion can be applied not only to the critical point
$(R_c, H_c(R_c))$, but to the entire line $H_c(R)$ at which the infinite
avalanche occurs.  In mean--field theory, the approach to this line is
continuous, with a power--law divergence of the susceptibility $\chi$
and precursor avalanches of all scales. Above 8 dimensions the action is
purely quadratic at the fixed point, and the infinite avalanche line
(where $1/\chi=0$ and $w=-2J^2\rho^{\prime}(-JM-H+k) \neq 0$)
presumably remains critical.
For $d=8-\tilde{\epsilon}$,
figure~\ref{Feynman}(a) shows the correction to vertex $w$ to first order in
$\tilde{\epsilon}$.
The incoming lines at a vertex stand for $\eta$ operators, and the
outgoing
lines are $\hat{\eta}$ operators.
The the low-frequency form of
the propagator is approximately $\delta(t-t^{\prime}))$ \cite{Narayan1}
but we have to observe causality; an example of a diagram forbidden
by causality is given in figure~\ref{Feynman}(b).
Applying the usual approximations
\cite{Goldenfeld}, we obtain for the recursion relation for $w$ to
$O(\epsilon)$:
\begin{equation}
w^{\prime} /2= b^{(-d/2+4)}\left\{w/2+(u_{2,0}/2) (w/2)^3 8 /(4 \pi)^4
\int_{\Lambda/b}^{\Lambda} dq \, 1/(q^2-\chi^{-1}/J)^4\right\}
\end{equation}
Writing
$b^{(-d/2+4)}=b^{(\tilde{\epsilon} /2)}=1+\tilde{\epsilon}/2 \,\log b$ and
performing the integral over the momentum shell $\Lambda/b<q<\Lambda$
leaves us with the recursion relation:
\begin{equation}
\label{w}
w^{\prime} /2=w/2+(w/2)\Bigl(\tilde\epsilon/2+u_{2,0}(w/2)^2 4 /(4
\pi)^4
\log b \Bigr)
\end{equation}
Since $u_{2,0}>0$ this means that for $\tilde\epsilon >0$ there are
only two fixed points with $w^{\prime}=w$: either $w=0$, which we will
discuss in the next paragraph, or $w=\infty$. We see that under the
recursion relation~(\ref{w}) any system that has a bare value $w
\neq 0$ when $1/\chi=0$ will flow to the fixed point $w=\infty$.
We interpret this
as indication that the transition is a first order transition
for $d<8$. Indeed, in three dimensions the simulation showed a
first-order transition without critical fluctuations for these systems.

The critical point we are interested in here is the fixed point where
$w = 0$. At $d=6$ the first non-quadratic contribution $u$ becomes
relevant,
{\it i.e.} the upper critical dimension~\cite{Maslov} for the critical
endpoint is 6.
We now compute, to $O(\epsilon)$, the corrections to the recursion
relations.
The relevant diagrams are shown in figure~\ref{Feynman}(c).
Figure~\ref{flows}(b) shows the corresponding RG flows in the
$(\chi^{-1},u)$ plane.

The loop-corrections look very similar to the loop-corrections in the
usual Ising model in $d-2$ dimensions.  In fact, to $O(\epsilon)$, they
are the same.  This can be seen either by direct computation (see
caption figure~(\ref{Feynman})), or by
noticing that the Feynman rules for our diagrams are the same as those
for the finite--temperature random-field Ising model\cite{Parisi}
(except that we have extra vertices which are irrelevant to
$O(\epsilon)$). This latter model has been mapped to all orders in
$\epsilon$ onto the regular Ising model, using supersymmetry and other
arguments. This analogy tells us that to $O(\epsilon)$ we get the same
RG flows (figure~1) and the same corrections to our exponents
as one finds in the usual Ising model in
$d-2$ dimensions\cite{Goldenfeld} (see table).

This mapping does not extend to the next $(\epsilon^2)$ term in the
series: figure~\ref{Feynman}(d) shows a correction of $O(\epsilon^2)$ to
the vertex $u_{2,2}$, which then contributes in $O(\epsilon^2)$ to the
propogator. This is comforting, as otherwise the critical properties of
our model in $d=3$ would have completely mapped onto the $d=1$ thermal
(non--random) Ising model, which has no finite--temperature phase
transition at all. Indeed, this was a substantive concern for the
thermal random--field Ising model, which despite the correspondence
above was proven to have a transition in $d=3$: the $\epsilon$-expansion
for that model summed over physically incorrect
metastable states.  By
controlling the history of the external field (as in
\cite{Narayan1,Narayan2}), we have been careful to specify the
particular metastable state in our calculations.

The $\epsilon$-expansion for our model is technically much simpler
than that for other disordered extended dynamical systems: {\it e.g.}
interface\cite{interface} or charge-density wave\cite{Narayan1} depinning,
where an infinite family of relevant operators made necessary a
functional renormalization group.  The relative ease of our calculation
may make possible further extensions: calculating the corrections
to the equations of state, calculating the history--dependent
critical behavior, or addressing the avalanche distributions.

We would like to thank O.~Narayan and D.~S. Fisher for advice and
consultation, and S.~Kartha, J.~A. Krumhansl, M.~E.~J. Newman, B.~W.~Roberts,
S.~Ramakrishna, J.~D. Shore, and J. von Delft
for helpful conversations, and
NORDITA where this project was started. We acknowledge the support of
DOE Grant \#DE-FG02-88-ER45364.

\begin{table}
\begin{tabular}{cr@{$\;=\;$}ld@{$\,\pm\,$}l}
exponents&
\multicolumn{2}{c}{$\epsilon$ expansion} &
\multicolumn{2}{c}{simulation\cite{hyster}}\\
&\multicolumn{2}{c}{with $\epsilon=d-6$, at $\epsilon = 3$} &
\multicolumn{2}{c}{in 3 dimensions}  \\ \hline
$1/\nu$ & $ 2 - \epsilon/3$&  1&1.0 & 0.1\\
$\beta $& $ 0.5 - \epsilon/6 $& 0&0.17 & 0.07\\
$\beta\delta$&$1.5 + O(\epsilon^2)$&1.5&2.0 & 0.3\\
$\delta$&$3+\epsilon$&6&\multicolumn{2}{c}{ (around 12)}\\
\end{tabular}

\caption{{\bf Universal exponents for critical behavior in
hysteresis loops.}
The exponents $\beta$ and $\delta$ tell how the magnetization scales
with $r$ and $h$, respectively, equation~(\protect\ref{magnetization}).
$\nu$ is the
correlation length exponent, measured (numerically) using finite--size
scaling.
}
\end{table}

 \begin{figure}
 \caption{{\bf Phase diagram and flows (schematic).} (1a) The vertical
axis is the
external field $H$, responsible for pulling the system from down to up.
The horizontal axis is the width of the random--field distribution $R$.
The bold line is $H_c(R)$, the location of the infinite avalanche (assuming
an initial condition with all spins down and a slowly increasing external
field).  The critical point we study is the end point of the infinite
avalanche line $(R_c, H_c(R_c))$.\hfill\break
\null\hskip0.2in
Using the analogy with the Ising model (see text) we also show the RG
flows around the critical point.  Here we ignore
the RG motion of the critical point itself: equivalently, the figure
can represent a section through the critical fixed point tangent to
the two unstable eigenvectors (labeled $h$ and $r$).  Two systems on
the same RG trajectory (dashed thin lines) have the same
long--wavelength
properties (correlation functions ...) except for an overall change in
length scale, leading to the scaling collapse of
equation~(\protect\ref{magnetization}).
The $r$ eigendirection to the left extends along the infinite avalanche
line; to the right, we speculate that it lies along the percolation threshold
for up spins (see reference\protect\cite{interface}).\hfill\break
\null\hskip0.2in
(1b) $O(\epsilon)$ RG flows below 6 dimensions in the
$(\chi^{-1},u)$ plane (see text). Linearization around the
Wilson-Fisher (WF) fixed point yields the exponents given to
$O(\epsilon)$ in the table. In the vicinity of the repulsive
$u=0=\chi^{-1}$ (MFT) fixed points one obtains the old mean-field
exponents.
 \label{flows}}
 \end{figure}

 \begin{figure}
 \caption{{\bf Feynman diagrams.} The perturbative expansion about mean--field
theory is presented here by Feynman diagrams.
(a) The correction to $O(\tilde{\epsilon})$ to the vertex w in an expansion
about 8 dimensions, see eq.~(\protect\ref{w}) in the text.
(b) An example of a diagram forbidden by causality.
(c) The relevant corrections
to first order in $\epsilon=6-d$ for the constant part $\chi^{-1}/J$
in the propagator and for $u$.
Using the same techniques
that lead to eq.~(\protect\ref{w}) we find the following recursion relations:
$(\chi^{-1}/J)^{\prime}=b^2\Bigl(\chi^{-1}/J+ u_{2,0} u
/(4 \pi)^3 \Lambda^2(1-1/b^2)/4 + u_{2,0} u /(4 \pi)^3 (\chi^{-1}/J) \log b
\Bigr) $ and
$u^{\prime}=u + u\Bigl(\epsilon+3 /(4 \pi)^3 u_{2,0} u\Bigr)\log b$.
$u_{2,0}$ does not get any loop corrections of $O(\epsilon)$.
(d) An example of a correction to the vertex $u_{2,2}$
which contributes only to $O(\epsilon^2)$, which is not present in the
regular Ising model (or the thermal random field Ising model).
 \label{Feynman}}
 \end{figure}

\end{document}